# A Novel Deep Learning Technique for Morphology Preserved Fetal ECG Extraction from Mother ECG using 1D-CycleGAN

Promit Basak[1⊕], A.H.M Nazmus Sakib[1⊕], Muhammad E. H. Chowdhury[2*], Nasser Al-Emadi[2], Huseyin Cagatay Yalcin[3], Shona Pedersen[4], Sakib Mahmud[2], Serkan Kiranyaz[2], Somaya Al-Maadeed[5]

[1]Department of Electrical and Electronic Engineering, University of Dhaka, Dhaka, 1000, Bangladesh.
 Email: basakpromit@gmail.com (P.B.); nazmussakib2970@gmail.com (N.S.)

[2]Department of Electrical Engineering, Qatar University, Doha, 2713, Qatar.
Email: mchowdhury@qu.edu.qa (M.E.H.C.); alemadin@qu.edu.qa (N.A.E.); sakib.mahmud@qu.edu.qa (S.M.); mkiranyaz@qu.edu.qa (S.K.)

[3]Biomedical Research Center, Qatar University, Doha, 2713, Qatar.
Email: hyalcin@qu.edu.qa (H.C.Y.)

[4]Department of Basic Medical Sciences, College of Medicine, Qatar University, Doha, 2713, Qatar.
Email: spedersen@qu.edu.qa (S.P.)

[5]Department of Computer Science and Engineering, Qatar University, Doha, 2713, Qatar.
Email: s_alali@qu.edu.qa (S.A.M.)

*Corresponding author: Muhammad E. H. Chowdhury (mchowdhury@qu.edu.qa, +974 31010775)
⊕ Both authors are equal contributors.

**Abstract**
Monitoring the electrical pulse of fetal heart through a non-invasive fetal electrocardiogram (fECG) can easily detect abnormalities in the developing heart to significantly reduce the infant mortality rate and post-natal complications. Due to the overlapping of maternal and fetal R-peaks, the low amplitude of the fECG, systematic and ambient noises, typical signal extraction methods, such as adaptive filters, independent component analysis, empirical mode decomposition, etc., are unable to produce satisfactory fECG. While some techniques can produce accurate QRS waves, they often ignore other important aspects of the ECG. Our approach, which is based on 1D CycleGAN, can reconstruct the fECG signal from the mECG signal while maintaining the morphology due to extensive preprocessing and appropriate framework. The performance of our solution was evaluated by combining two available datasets from Physionet, "Abdominal and Direct Fetal ECG Database" and "Fetal electrocardiograms, direct and abdominal with reference heartbeat annotations", where it achieved an average PCC and Spectral-Correlation score of 88.4% and 89.4%, respectively. It detects the fQRS of the signal with accuracy, precision, recall and F1 score of 92.6%, 97.6%, 94.8% and 96.4%, respectively. It can also accurately produce the estimation of fetal heart rate and R-R interval with an error of 0.25% and 0.27%, respectively. The main contribution of our work is that, unlike similar studies, it can retain the morphology of the ECG signal with high fidelity. The accuracy of our solution for fetal heart rate and R-R interval length is comparable to existing state-of-the-art techniques. This makes it a highly effective tool for early diagnosis of fetal heart diseases and regular health checkups of the fetus.

## 1. Introduction

ECG signal analysis is a common technique for monitoring and diagnosing a range of common heart conditions (Jeffries, 2003). The signal is obtained by placing electrodes on an adult's chest, hands, or legs (Hasan, 2007). This technique can also be adapted to acquire a fetal electrocardiogram (fECG), which, among other benefits, enables the detection of fetal heart abnormalities. Procedures for fECG recording can be invasive or non-invasive. Invasive fECG recording involves placing electrodes on the fetal scalp only at the final stage of pregnancy. The invasive method for fECG recording produces excellent results but can cause several complications and infections (Barnova, 2021b). However, clinical care and delivery planning can benefit from early diagnosis of aberrant fECG at earlier stages of pregnancy. A non-invasive alternative is to carry out a mother ECG and separate the fECG from the mother Electrocardiogram (mECG), which is recorded by placing the electrodes on the mother's abdomen. Non-invasive fECG (NI-fECG) is a promising diagnostic method that can be used to diagnose a variety of fetal cardiac conditions early in pregnancy. These include arrhythmias, septal defects, and aortic stenosis, as well as many similar genetic conditions which can be diagnosed by measuring ECG parameters such as fetal heart rate (FHR), heart rate variability (HRV) and ECG morphological information (PR, ST and QT intervals) (Clifford, 2014) etc. For example, the FHR is calculated from the recorded fECG and its value is compared to the normal heart rate for the corresponding gestational age. A normal heart rate indicates that the mother and fetus are receiving an adequate amount of oxygen. On the other hand, anomalies in heart rate may indicate a problem with the fetal cardiac system or the oxygen transfer route (Anisha, 2021).

In addition to NI-fECG, there are several other non-invasive methods for fetal heart rhythm monitoring. These methods are useful for obtaining electrical stimulation waveforms similar to the ECG recordings. In ultrasonic fetal cardiotocography (CTG), pressure transducers are used that detect uterine contraction to track fetal heart rates. However, CTG cannot measure beat-to-beat heart rate data and there are potential safety concerns associated with exposure to ultrasound irradiation, which makes this method inappropriate for long-term monitoring (Peters, 2001). Fetal magnetocardiography (FMCG) is another reliable technique that uses SQUID (Superconductive Quantum Interference Device) sensors placed close to the mother's abdomen to detect the magnetic field of the fetal heart from which electrical cardiac waveforms can be extracted. Despite providing high-quality signals, FMCG is far less popular due to its cost and complicated installation. Comparing these methods, NI-fECG is inexpensive, practical, user-friendly, and non-invasive and at the same time, makes it possible to monitor and morphologically analyze the fetus's heart continuously, even during labor (Clifford, 2014).

Extraction of fECG from mECG is rather challenging as mECG often contains a variety of noises resulting from a variety of sources such as baseline drift, motion artifacts, power-line noise, uterine and muscle contractions, loose electrode connection, and white noise (Clifford, 2006; Hasan, 2007). This makes fECG extraction very challenging, even if only a slight amount of distortion is present in the mECG. A typical fECG has several significant elements like P waves, ST segments, and T waves which are seen to be useful in diagnosing many diseases in the fetal stage (Shepoval'nikov, 2006). Still, most of the previous works on abdominal mECG focused on detecting only the QRS complex, not the whole fECG. Several algorithms and techniques including adaptive filters, least mean square (LMS), recursive least square (RLS), Kalman filters and extended state Kalman filters are used in fECG extraction problem (Ferrara, 1982; Niknazar, 2012; Rafaely, 2000). However, these algorithms cannot work efficiently when mECG and fECG R-peaks coincide and require a high signal to noise ratio (SNR) value. On the other hand, independent component analysis (ICA) and principal component analysis (PCA)-based techniques are not robust enough and require specific electrode arrangements (Behar, 2014; Martinek, 2018). The issues with the aforementioned algorithms make the task of extracting fECG from mECG even more challenging.

Because of the significant drawbacks of traditional methods, *deep learning*-based approaches can be utilized. According to W.J. Zhang et al. (W. Zhang, 2018), the outcome of learning or deep learning is a mapping from inputs to outputs (a class or an instance). Compared to traditional techniques, deep learning can be used to learn very complex mappings which can be used to suppress the mECG to extract only the fECG. Generative Adversarial Networks (GANs) show excellent performance in noise reduction (Chen, 2020; Tran, 2020), which can be used to suppress noise and ensure efficacious translation between input and label signals. Specifically, CycleGAN has recently been used in a variety of different paired generation tasks including non-parallel voice conversion (Kaneko, 2019), X-ray style transfer (Tmenova, 2019) and MR to CT conversion (Yang, 2018). It is reported that CycleGAN can learn to hide or suppress any particular information or signal making it the perfect choice for our task (Chu, 2017).

Our solution using 1D CycleGAN with the proposed loss function is able to provide a reconstruction of fECG that is unaffected by human errors. The contributions of this study can be described as follows:

- While most existing solutions focus on fQRS detection, our solution can reconstruct the whole fECG signal while preserving its morphology. This means that information about P, S, T waves and PR, ST and QT intervals are obtainable from the reconstructed signal, which is essential for diagnosing various diseases.
- It provides comparable performance to the state-of-the-art techniques in case of fQRS detection, which is useful for R-peak detection.
- Our approach uses a novel weighted loss that significantly improves the quality of the generated fECG signal.
- Our preprocessing module includes robust techniques that discard artifacts i.e., filter lag, transient response, etc.
- It can measure fetal heart rate and heart rate variability metrics with great accuracy from the generated fECG signal.
- We used a mixture of two different real-world datasets which makes the framework more robust and independent of the experimental setup, electrode position, recording equipment and other biases.

This document has been structured into five sections. In section 2, we discuss related works with their pros and cons. Then we provide a comprehensive description of the datasets and the conceptual framework of our proposed methodology in the next section. In section 4, the findings of this study are summarized. Finally, the concluding discussion and future prospects are given in section 5.

## 2. Related Works

Although the majority of earlier research has concentrated on the extraction of QRS waves (Varanini, 2017; Zhong, 2018), the entire fECG signal waveform must be extracted for a complete assessment (Shepoval'nikov, 2006). The use of adaptive filters can cause the removal of some parts of the fECG signal as filters' coefficients change in response to a time-varying signal, which presents various difficulties for mECG signal processing. First of all, the power spectral density (PSD) of the input signal influences how quickly adaptive filter algorithms converge (Rafaely, 2000). The least mean-square error can be used as the objective function to achieve convergence, but adaptive filters require a constant and flat power spectrum, which does not match the power spectrums of real-world signals. Second, colored noise components in real-world data significantly reduce the efficiency of adaptive filters (Mumford, 2010).

On the other hand, least mean square (LMS) and recursive least square (RLS) algorithms are typically employed for narrowband frequencies (Rafaely, 2000). However, they require a reference signal closely resembling the morphology of the mECG waveform. Therefore, these algorithms are optimized using Weiner's optimal solution. When the peaks of the mECG and fECG coincide, techniques that rely on temporal characteristics, such as template-based and conventional Kalman filters, may also be deemed ineffective. For reliable fECG extraction, the extended state Kalman filter was developed, which may address the QRS coincidence issue (Niknazar, 2012). However, due to their higher computational cost, they are not very effective and are unable to detect R-peaks precisely. Different forms of blind source separation methods, such as independent component analysis (ICA) and principal component analysis (PCA), are available (Behar, 2014) to address the drawbacks of adaptive filters. These techniques are built using various abdominal channels to create a linear stationary mixing matrix (Martinek, 2018). One study used a blind source separation technique to identify a reference signal for the adaptive filter to extract fECG from a single channel, achieving a 96% F1 score for fetal QRS (fQRS) detection. However, these techniques have a specific electrode configuration and a low signal-to-noise ratio (SNR). In addition, they necessitate intensive post-processing for higher caliber fECG extraction (Mohebbian, 2020).

Jezewski et al. (Jezewski, 2012) determined fetal heart rate from abdominal ECG signal after the detection of fetal QRS complex and compared the results with Doppler ultrasound monitoring techniques. The study recognizes indirect ECG-based fetal heart rate detection to be better than ultrasound monitoring techniques. On the downside, the process of inference was handcrafted and offline which is not suitable for real-life applications. Jaros et al. (Jaros, 2019) used a combination of independent component analysis (ICA), adaptive neuro-fuzzy inference system (ANFIS) algorithm and recursive least square (RLS) algorithm for fECG extraction. However, these techniques suffer when there is an overlap between mother QRS and fetal QRS.

Zhang et al. (N. Zhang, 2017) achieved a 99% F1 score for QRS complex recognition by combining smooth window and singular value decomposition (SVD). In addition to proposing the prefix tree-based method known as QRStree for QRS detection, Zhong et al. (Zhong, 2018) also suggested a convolutional neural network (CNN) for QRS complex detection and achieved a 77% F1 score. Additionally, this study proposed a prefix tree-based method named QRStree for QRS detection and achieved a 95% F1 score. This study uses a string of alphabetical letters to represent sequential fQRS. The strings were stored within a prefix tree, and an ideal path was selected for precise fQRS detection by taking advantage of the connections made by fQRS in the tree.

Mohebbian et al. (Mohebbian, 2021) used attention-based CycleGAN to extract fECG and achieved 99.7% F1-score [CI: 95%: 97.8-99.9], 99.6% F1-score [CI: 95%: 98.2%, 99.9%] and 99.3% F1-score [CI: 95%: 95.3%, 99.9%] for fetal QRS detection on the A&D fECG, NI-fECG and NI-fECG challenge datasets, respectively. They discovered that convolutional kernels tend to amplify both the maternal and the fetal R waves, which is undesirable. To counteract this, an attention layer was implemented that masked specific parts of the signal, as processing those parts would increase error. Additionally, they used 1D convolutional layers with a sine activation function. They deemed it superior to popular options such as Leaky ReLU because it tends to preserve more signal detail and has demonstrated favorable results in representing audio, video, and image signals.

In many of these studies, QRS waves are accurately identified; however, other components of the fECG signal must be considered. A universal model that could be applied to participants with diverse electrode locations and surroundings was also not examined in earlier studies. The F1 score was used to evaluate performance in practically all of the published work for QRS detection, although signal reconstruction metrics like mean absolute error (MAE), root mean squared error (RMSE), and correlation coefficient (CC) are not provided.

## 3. Methodology

In this section, we will commence with an overview of our methodology. After discussing the datasets used in this study, we will explain the proposed methodology in detail.

*3.1 Overview*

The general framework of our proposed methodology is given in **Figure 1**. A four-channel mECG is collected by placing four electrodes on the abdomen of the mother. The acquired signal has three components: pure mECG, fECG, and noise.

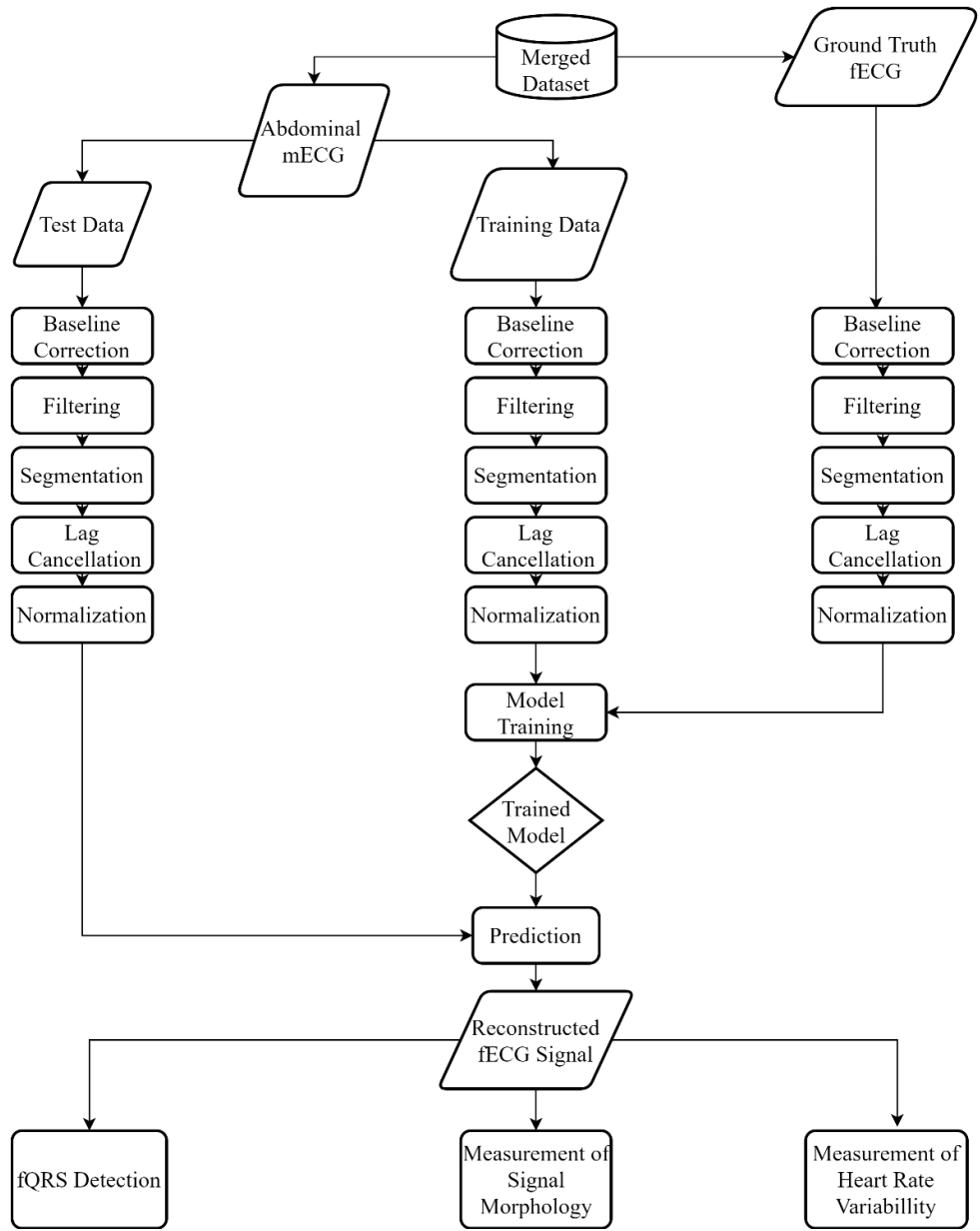

**Figure 1**: General framework of our proposed methodology.

fECG is recorded by connecting a single electrode to the scalp of the fetus. Our goal is to extract this fECG from the four-channel mECG. Initially, the signals undergo a few preprocessing steps, such as resampling, baseline correction, filtering, segmentation, and normalization. Different variants of the 1D CycleGAN are trained using the fECG signals collected from the fetal scalp as ground truth. Afterwards, the trained models are used to generate predictions from the test signals, which go through the same preprocessing steps. We then choose the optimal model by evaluating the generated signals using Pearson Cross-Correlation (PCC), Spectral Correlation, Spectral RMSE, MAE, MAPE, and RMSE as evaluation metrics. Further elaborated information about the datasets and the methodology are provided in the subsequent sections.

*3.2 Dataset Description*
*3.2.1 Dataset 1*
The first dataset we used is the "Abdominal and Direct fECG (A&D fECG) Database" from Physionet ([Jezewski, 2012](Jezewski, 2012)). This dataset includes 4 channel mECG signals that were recorded from 5 women's abdomens between 38 and 41 weeks of pregnancy. Additionally, it includes the simultaneous fECG recorded from the fetus' scalp and expert-annotated R-peak locations. With the help of four abdominal electrodes positioned around the navel, the 4-channel ECG signal was recorded (**Figure 2**). A reference electrode was attached above the pubic symphysis, and a common mode reference electrode was positioned on the left leg. These electrodes are placed in the exact location throughout the experiment. To increase skin conductivity, abrasive material was employed along with Ag-AgCl electrodes. The corresponding experimental setup is shown in **Figure 2**.

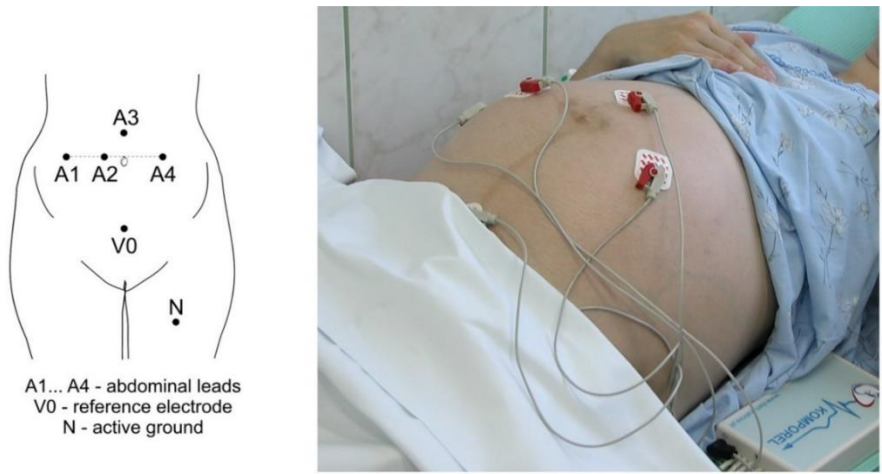

**Figure 2**: Experimental setup (Matonia, 2020).

Five records for five subjects are included in the data set. Each recording contains a comparable 5-minute single-channel fECG and 4-channel mECG signals. All signals were sampled with a 16-bit resolution at 1000 Hz. The bandwidth of the signal is 1-150 Hz. The sample signal segment for the four-channel mECG and ground truth fECG in Dataset 1 is shown in **Figure 3**.

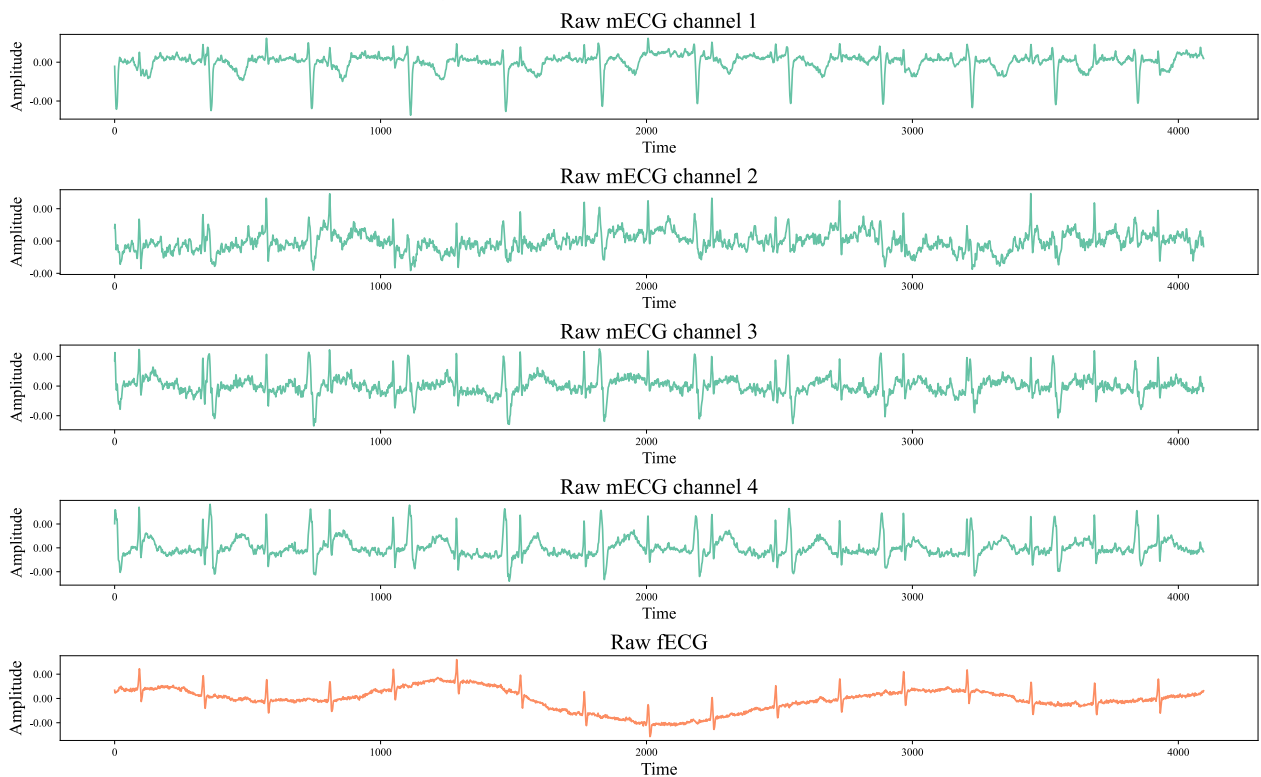

**Figure 3**: Raw data from dataset 1.

*3.2.2 Dataset 2*

The second dataset used in this study is "Fetal electrocardiograms, direct and abdominal with reference heartbeat annotations" from Matonia et al. (Goldberger, 2000; Matonia, 2020). This dataset is divided into two sections: a) labor signals and b) pregnancy signals. In the remainder of the paper, we refer to labor signals as "Dataset 2a" and pregnancy signals as "Dataset 2b". 4 channel mECG signals from the abdomen of 12 women were recorded for Dataset 2a, 5 of these during the late stages of childbirth where 500 Hz is used for sampling the mECG. There are 12 records for 12 women, and each record has an ECG signal that lasts for five minutes.

Additionally, the simultaneous fECG taken from the scalp of the fetus and expert-annotated fECG R-peak sites are included in dataset 2a. The sample rate for the fECG signal is 1000 Hz.

In Dataset 2b, ten women's four channels abdominal mECGs from the 32nd and 42nd weeks of gestation were recorded. The mECG is sampled at 500 Hz. Each record has a 20-minute signal and contains 10 records for 10 women. The concurrent fECG recorded from the fetus's scalp is not included. Instead, it has expert-labeled R-peak positions of fECG. A sample of raw data from dataset 2 is shown in **Figure 4**.

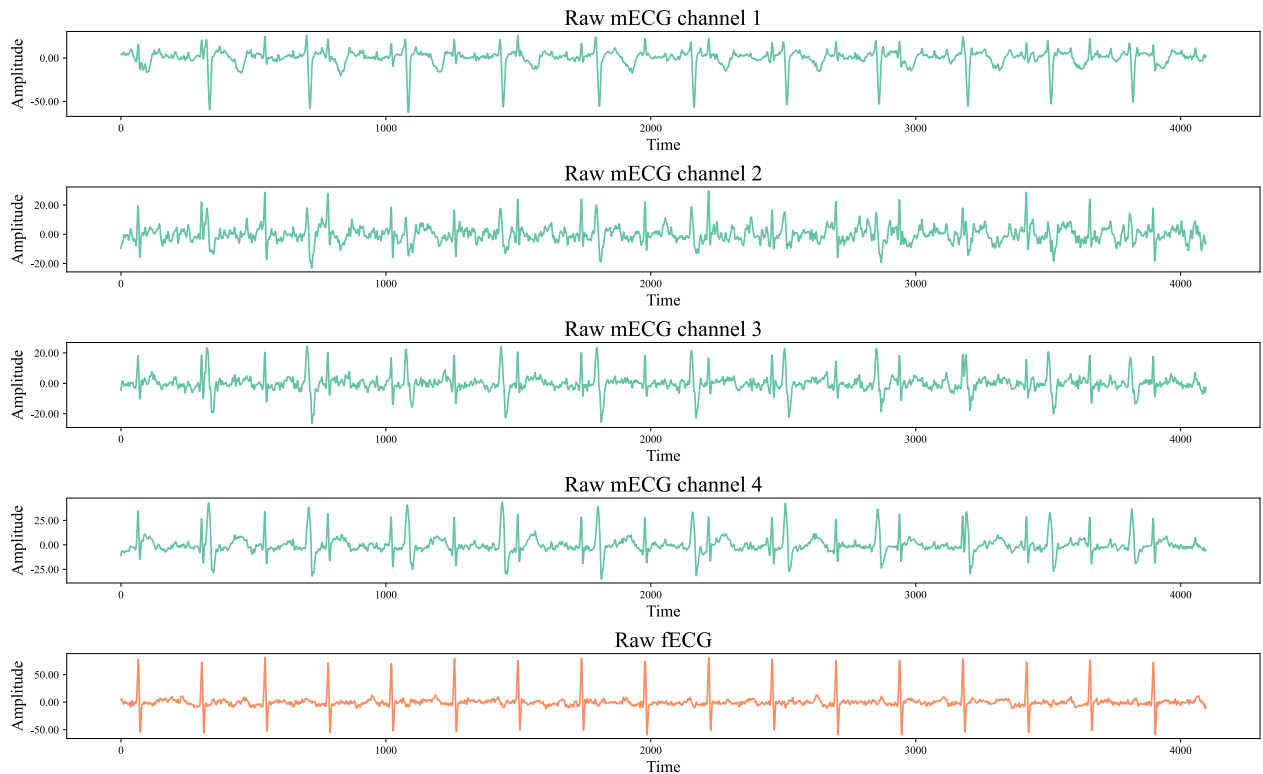

**Figure 4**: Raw data from dataset 2.

### 3.3 Proposed Method
#### 3.3.1 Preprocessing

Biomedical signals are corrupted by numerous noises and distortions, including but not restricted to baseline drift, motion artifacts, and power line noise ([Hossain, 2022](#)), ([Kiranyaz, 2022](#)), ([Shuzan, 2021](#)). In the datasets, both the fECG and mECG are harmed severely by these disturbances. To address these issues, several effective preprocessing techniques are chosen carefully and applied together with a sophisticated strategy to retain the signal's morphology as much as possible. The fECG and the mECG signals from both datasets are first resampled to 512 Hz to maintain uniformity in sampling frequency. This sampling frequency is determined empirically through testing. A bandstop filter is applied to both the mECG and fECG signals with a center frequency of approximately 50 Hz to remove the power line noise. According to the recommendation by Bailey et al., the minimum bandwidth of a clinical ECG should be 75 Hz to 100 Hz ([Bailey, 1990](#)). Moreover, the QRS complex of an fECG signal lies in a range of 10 to 15 Hz, and the dominant frequency of an ECG is under 35 Hz ([Sameni, 2010](#)). Therefore, a higher-order Butterworth bandpass filter is applied with cutoff frequencies of 0.1 Hz to 70 Hz. Both the mECG and fECG signals suffer from the baseline drift issue. To solve this problem, an approach is adopted where a polynomial is fitted to the baseline drift and subtracted from the signal subsequently. This technique is applied to the mECG and fECG signals with orders 8 and 36, respectively, which are selected through trial and error. After this step, moving average filters of window sizes 4 and 10 are applied to the mECG and fECG, respectively, to smooth out the signals. Then the data is segmented with a window size of 512 (1 second), and baseline fixing is applied to the individual segments to remove the local baseline drift. Each signal segment is then range normalized between 0 and 1 using the following formula:

$$X_n(i) = \frac{X(i) - X_{min}}{X_{max} - X_{min}} \quad (1)$$

Here, $X(i)$ is the original amplitude of the segment, $X_n(i)$ is the normalized segment, $X_{min}$ and $X_{max}$ are the minimum and maximum amplitude of the signal, respectively.

The final signal matrix shape is $N \times M \times C$, where $N$ is the total number of signal segments, $M$ is the length of the signal, and $C$ is the total number of channels. In this study, $C$ is 4 and 1 for mECG and fECG, respectively. A comparison between the raw and preprocessed data is shown in **Figure 5.**

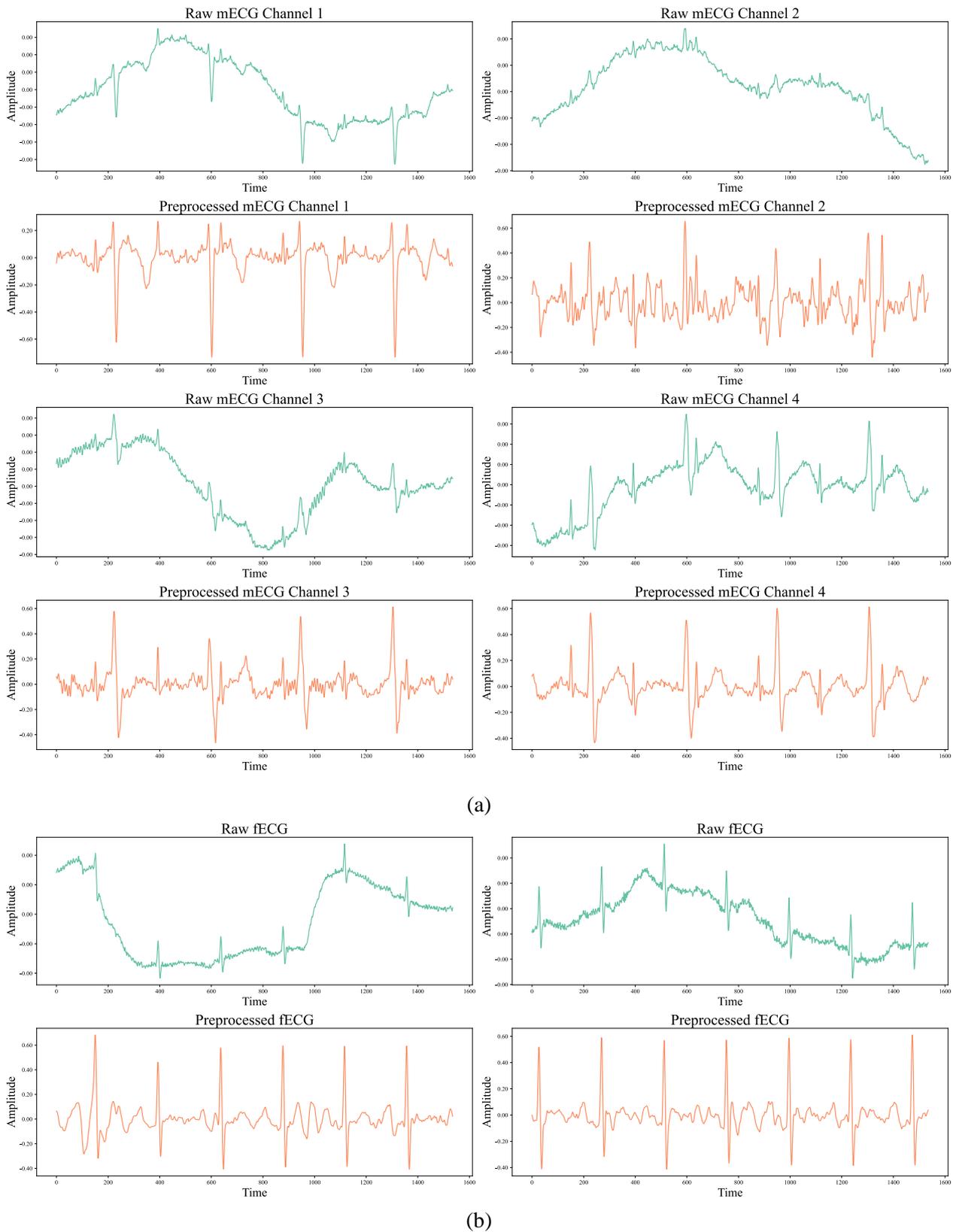

**Figure 5**: Comparison between raw fECG and preprocessed fECG (a), 4-channel raw mECG and preprocessed mECG (b) (top: raw, bottom: preprocessed).

There are a few complications while applying the preprocessing techniques to the ECG signals. In this paper, higher-order filters are chosen because of their ability to provide more attenuation and a narrower transition band, resulting in better separation between the passband and the stopband. However, higher-order filters often cause phase lag, which is completely unwanted and destroys the purpose of preprocessing. To eliminate this phase lag, an apply-flip-reapply technique is employed. First, the higher-order filter is applied to the signal. After that, the signal is flipped temporally, and the same filter with identical parameters is applied again to the flipped version, which cancels the phase lag. Finally, the signal is flipped again to get the original signal with zero lag. The lag cancellation process is described in **Figure 6**. Another problem is the presence of ripples at the start and end of the signal after baseline correction. This problem is addressed by applying an overlapping window and slicing the ripples from the beginning and the end.

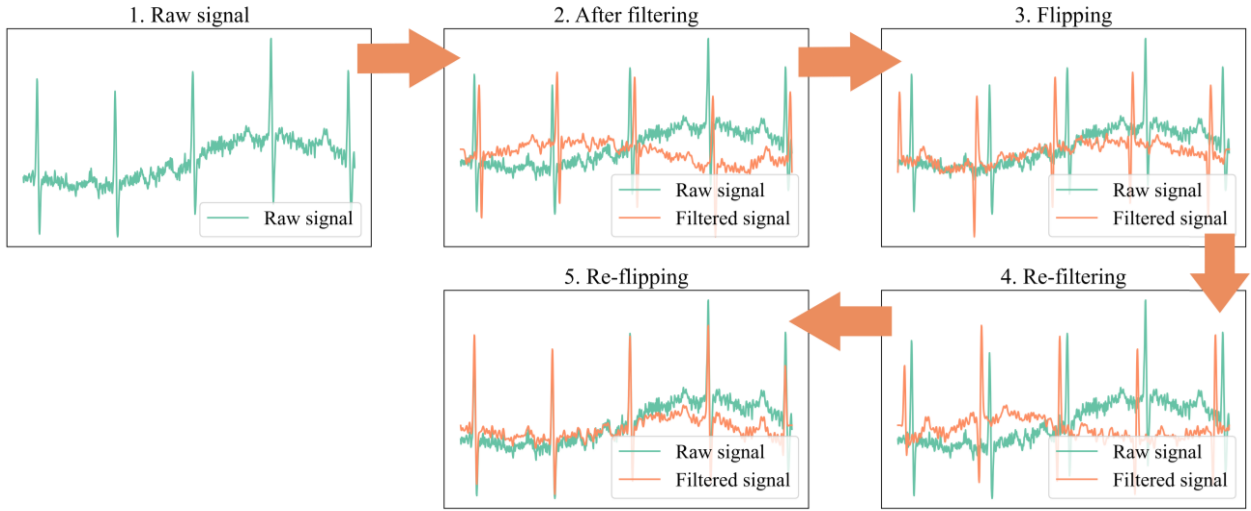

**Figure 6**: Lag Cancellation of the higher order filter.

*3.3.2 Model Architecture*

Here, we first describe the architecture of the main framework, losses, generator and then the discriminator of the proposed methodology.

*3.3.2.1 CycleGAN*

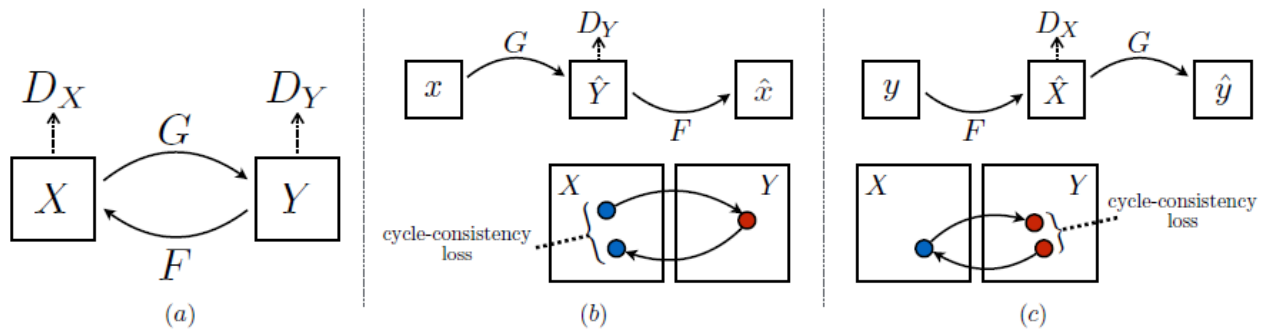

**Figure 7**: Original CycleGAN architecture (Zhu, 2017).

In this study, we propose a 1D CycleGAN-based model which extracts the fECG from a four-channel mECG. CycleGAN, as described in **Figure 7**, is essentially a paired network that learns two mappings using two generators: $G_1: X \rightarrow Y$ and $G_2: Y \rightarrow X$. For each mapping, a Generative adversarial network (GAN) (Goodfellow, 2020) network is trained such that $\hat{y} = G_1(X)$ is almost identical to $Y$ and $\hat{x} = G_2(Y)$ is very much similar to $X$. Mathematically, $G_1$ and $G_2$ should be inverse of each other and to maintain the consistency of these two corresponding networks, we add a cycle consistency loss (Zhou, 2016), which helps to maintain the properties $G_1(G_2(Y)) = Y$ and $G_2(G_1(X)) = X$ (Zhu, 2017). Two adversarial discriminators, $D_1$ and $D_2$, are associated with the generators of the GAN networks, which compute the similarity between the ground truth and generated output of the mapping. Hence, discriminators need another type of loss and we call this second one adversarial loss or GAN loss. Finally, the discriminators are connected to the alternate generators to complete the cycle. In the original implementation, adversarial loss ($\mathcal{L}_{adv}$) and cycle consistency loss ($\mathcal{L}_{cyc}$) are defined as follows:

$$\mathcal{L}_{adv}(G_1, X, Y, D_2) = \frac{1}{N}\sum_{i=1}^{N} y_i[\log D_2(y_i)] + x_i[1 - \log D_2(G_1(x_i))] \qquad (2)$$

$$\mathcal{L}_{cyc}(G_1, G_2) = \frac{1}{N}\sum_{i=1}^{N}\|G_2(G_1(x_i)) - x_i\|_1 + \frac{1}{N}\sum_{i=1}^{N}\|G_1(G_2(y_i)) - y_i\|_1 \qquad (3)$$

Here, $N$ signifies the total number of training examples and $\|x\|_1$ denotes the L1 norm of $x$, $G_1$ and $G_2$ are generators of the CycleGAN, $D_1$ and $D_2$ are corresponding discriminators.

In this paper, we want to map mECG signals to fECG signals. Hence, we need one generator mapping mECG to fECG ($G_1$) and another mapping fECG to mECG ($G_2$). Their corresponding discriminators are denoted as $D_1$ and $D_2$. Instead of log loss, we initially used the $L1$ loss ($\mathcal{L}_{L1}$) as the adversarial loss. However, in the case of ECG, it is also critical to preserve the morphology, position of QRS complex and spectral components. So, we added three more loss components to the adversarial loss: spectral loss ($\mathcal{L}_{spec}$), temporal loss ($\mathcal{L}_{temp}$) and power loss ($\mathcal{L}_{power}$) to make the extracted signal morphologically very close to the actual ECG signal.

$$\mathcal{L}_{L1} = [1 - D_1(G_1(X))]^2 + [1 - D_2(G_2(Y))]^2 \qquad (4)$$

$$\mathcal{L}_{spec} = \frac{1 - \rho(PSD(Y), PSD(G_1(X)))}{\rho(PSD(Y), PSD(X))} \qquad (5)$$

$$\mathcal{L}_{temp} = 1 - \rho(Y, G_1(Y)) \qquad (6)$$

$$\mathcal{L}_{power} = \left|\frac{Power_Y - Power_{G_1(X)}}{Power_Y}\right| \qquad (7)$$

Here, $PSD$ signifies the power spectral density, $Power_Y$ is the power of the signal of interest and $\rho$ is the Pearson's correlation coefficient (PCC). $PCC$ is a measure of correlation between two signals and can be formulated by:

$$\rho(x, y) = \frac{\sum_{i=1}^{N}(x_i - \bar{x})(y_i - \bar{y})}{\sqrt{\sum_{i=1}^{N}(x_i - \bar{x})^2}\sqrt{\sum_{i=1}^{N}(y_i - \bar{y})^2}} \qquad (8)$$

Here, $\bar{x}$ and $\bar{y}$ are the mean values of signals $x$ and $y$ respectively and $N$ is their length.

Thus, the modified adversarial loss became:

$$\mathcal{L}_{adv} = \mathcal{L}_{L1} + p \times \mathcal{L}_{spec} + q \times \mathcal{L}_{temp} + r \times \mathcal{L}_{power} \qquad (9)$$

Here, $p$, $q$ and $r$ are weight coefficients for spectral loss, temporal loss and power loss accordingly. Experimentally, we found $p = 2$, $q = 4$, and $r = 1$ to produce the best ECG signals. These three weighted loss components played a very important role in extracting fECG signals which are described in ablation studies.

A general-purpose training algorithm of the proposed CycleGAN architecture is provided below:

---

**Algorithm 1**: Training algorithm for CycleGAN for mini-batch gradient descent.
$k$ is the parameter for mini-batch number.

---

**for** number of training iterations **do**

    **for** $k$ steps **do**

- Draw a minibatch of $m$ samples $\{x^{(1)}, \ldots, x^{(m)}\}$ from domain X
- Draw a minibatch of $m$ samples $\{y^{(1)}, \ldots, y^{(m)}\}$ from domain Y
- Calculate the discriminator loss on the ground truth signals:

$$J_{D_{ground}} = \frac{1}{m}\sum_{i=1}^{m}(D_X(x^{(i)}) - 1)^2 + (D_Y(y^{(i)}) - 1)^2$$

- Calculate the discriminator loss on the target signals:

$$J_{D_{target}} = \frac{1}{m}\sum_{i=1}^{m}\left(D_Y\left(G_X(x^{(i)})\right)\right)^2 + \left(D_X\left(G_Y(y^{(i)})\right)\right)^2$$

- Update the discriminators.
- Calculate the generator $G_Y$ loss:

$$J_{cycle}^{(G_Y),(G_X)} = \frac{1}{m}\sum_{i=1}^{m}\|G_X(G_Y(y_i)) - y_i\|_1$$

$$J_{spec_Y} = \frac{1}{m}\sum_{i=1}^{m}\frac{1 - \rho(PSD(x^{(i)}), PSD(G_Y(y^{(i)})))}{\rho(PSD(x^{(i)}), PSD(y^{(i)}))}$$

$$J_{temp_Y} = \frac{1}{m}\sum_{i=1}^{m}1 - \rho\left(x^{(i)}, G_Y(y^{(i)})\right)$$

$$J_{power_Y} = \frac{1}{m}\sum_{i=1}^{m}\left|\frac{Power(y^{(i)}) - Power(G_Y(y^{(i)}))}{Power(y^{(i)})}\right|$$

$$J_{G_X} = \frac{1}{m}\sum_{i=1}^{m}\left(D_X(G_Y(y^{(i)})) - 1\right)^2 + J_{cycle}^{(G_Y),(G_X)} + p \times J_{spec} + p \times J_{temp} + p \times J_{power}$$

- Calculate the generator $G_X$ loss:

$$J_{cycle}^{(G_X),(G_Y)} = \frac{1}{m}\sum_{i=1}^{m}\left\|G_Y(G_X(x^{(i)})) - x^{(i)}\right\|_1$$

$$J_{spec_X} = \frac{1}{m}\sum_{i=1}^{m}\frac{1 - \rho\left(PSD(y^{(i)}), PSD(G_X(x^{(i)}))\right)}{\rho\left(PSD(x^{(i)}), PSD(y^{(i)})\right)}$$

$$J_{temp_X} = \frac{1}{m}\sum_{i=1}^{m} 1 - \rho\left(y^{(i)}, G_x(x^{(i)})\right)$$

$$J_{power_X} = \frac{1}{m}\sum_{i=1}^{m}\left|\frac{Power(x^{(i)}) - Power(G_X(x^{(i)}))}{Power(x^{(i)})}\right|$$

$$J_{G_X} = \frac{1}{m}\sum_{i=1}^{m}\left(D_Y(G_X(x^{(i)})) - 1\right)^2 + J_{cycle}^{(G_X),(G_Y)} + p \times J_{spec} + p \times J_{temp} + p \times J_{power}$$

- Update the generators.

**endfor**

**endfor**

*3.3.2.2 Generator*

One of the most prominent issues of deep neural networks is the degradation problem, where the accuracy saturates and degrades rapidly as the number of layers increases. To address this issue, He et al. (He, 2016) proposed a solution to this problem by introducing a deep residual learning framework popularly known as Resnet. Every few stacked layers, residual learning is adopted. A building block (shown in **Figure 8(a)**) is defined as:

$$y = F(x, \{W_i\}) + x \quad (10)$$

Here, the input and output vectors for the layers under consideration are $x$ and $y$. The residual mapping to be learned from input $x$ and weights $W_i$ is represented by the function $F(x, \{W_i\})$. By using a shortcut connection and element-wise addition, the operation $F + x$ is carried out. This shortcut connection does not introduce any additional parameters or computational complexity. If the dimensions of $x$ and $F$ are not equal, a linear projection $W_{sx}$ is performed by the shortcut connections to match the dimensions:

$$y = F(x, \{W_i\}) + W_{sx} \quad (11)$$

The form of the residual function $F$ is flexible as any number of layers can be implemented per stack. The notations presented here of these equations are for fully-connected layers for simplicity, but they are also applicable to convolutional layers. The element-wise addition is performed channel by channel.

The generator of our proposed architecture is divided into downsampling and upsampling layers, with 'n' Resnet blocks described beforehand in between. Our Resnet blocks contain two units, each of which contains a ReflectionPad2d layer followed by 64 Conv1d layers and a batch normalization layer, with a fully connected and a dropout layer in between. A cross-connection is made from the end of the second unit to the beginning of the first unit.

The generator takes as input 4-channel mECG data. These data are padded and passed through three one-dimensional convolutional layers of sizes 16, 32 and 64. Data are batch normalized and pass through a dense layer with the 'ReLU' activation function after every convolutional layer. These layers together are the downsampling layers.

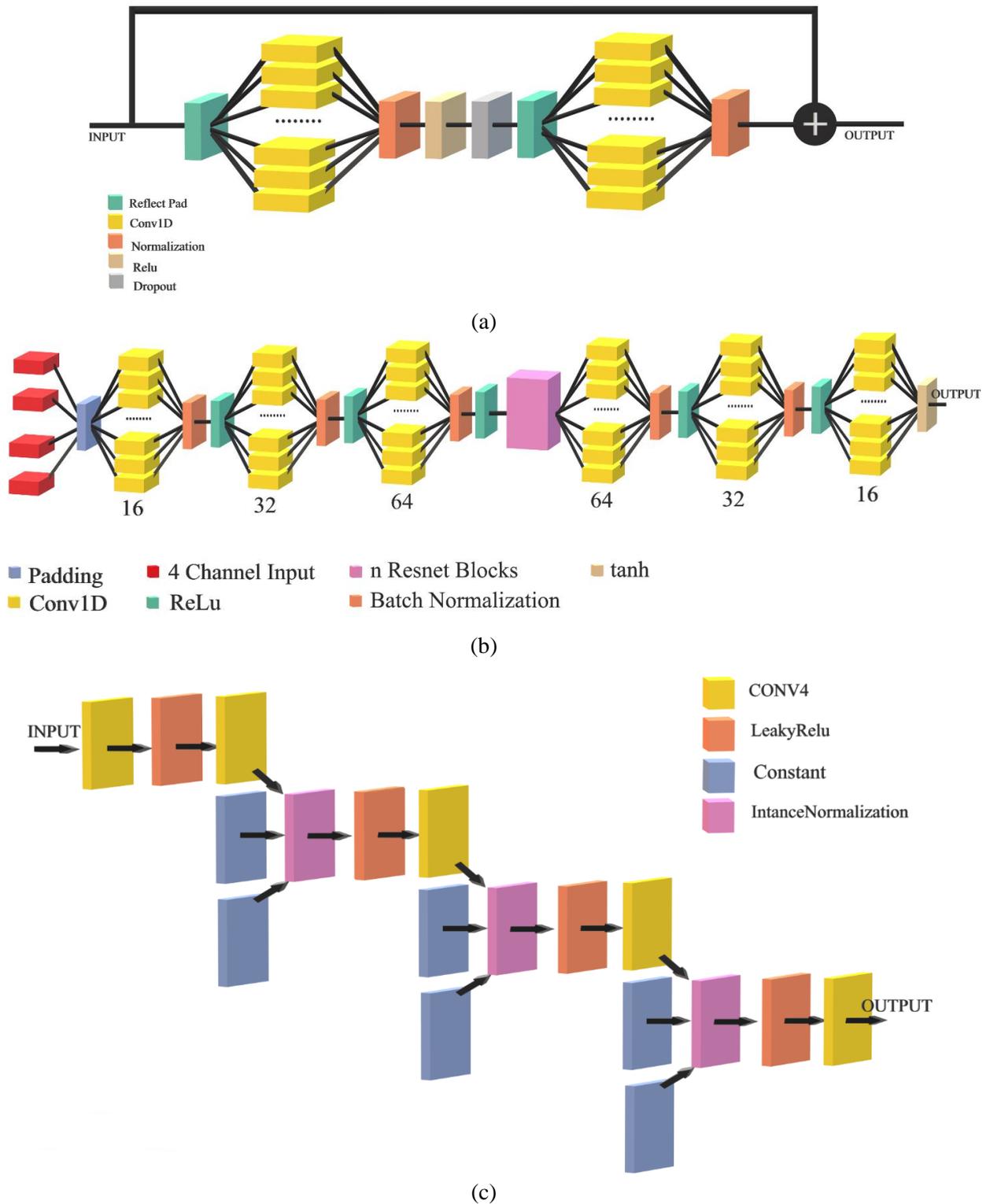

**Figure 8:** Architecture of the Proposed model: (a) the Resnet Block, (b) the Generator Block, and (c) the Discriminator Block.

After the data has passed through the Resnet blocks, we upsample the data using transposed convolution, where it passed through three ConvTranspose1D layers of sizes 16, 32 and 64, consecutively. After the first two layers, the data are batch normalized and passed through a dense layer with the 'ReLU' activation function. After the third ConvTranspose1D layer, the data are passed through a dense layer with the 'tanh' activation function. The complete architecture of the generator is represented pictorially in **Figure 8(b)**.

We have investigated different generators, including variants of ResNet, Unet and Feature Pyramid Network (FPN). It is worth noting here that the non-linear attributes of the generative neuron can allow the model to learn in a compact architecture. Operational Neural Network (ONN) utilizes such generative neuron which was optimized to create a self-organized ONN model or Self-ONN model. As discriminators, we used "Basic Discriminator" and "Self-Discriminator", the basic discriminator being a $3 \times 3$ PatchGAN described in the methodology section. Self-discriminator is essentially a $1 \times 1$ PatchGAN with 1D Self-ONN layers. Similarly, the FPN used in the study was not a vanIlla FPN model rather it was a Self-FPN model.

*3.3.2.3 Discriminator*

We used a $3 \times 3$ PatchGAN as the discriminator of the CycleGAN architecture depicted in **Figure 8(c)**. PatchGAN solely penalizes structure at the scale of local picture patches. It is a sort of discriminator for generative adversarial networks. Each patch of an image is evaluated by the PatchGAN discriminator to determine if it is real or fake. Convolutionally applied to the image, this discriminator produces an output by averaging all responses. With the assumption of independence between pixels separated by more than a patch diameter, such a discriminator effectively models the image as a Markov random field. It may be interpreted as a certain texture or style loss (Isola, 2017).

One kind of GAN that utilizes labels during training is a conditional generative adversarial network (CGAN). Its objective can be expressed as (Isola, 2017):

$$\mathcal{L}_{cGAN}(G,D) = \mathbb{E}_{x,y}[\log D(x,y)] + \mathbb{E}_{x,z}[\log 1 - D(x, G(x,z))] \quad (12)$$

where $\mathbb{E}$ signifies the expected value of the sample.

Oftentimes, a more traditional loss is mixed with the GAN's objective. In this case, the L1 loss is added.

$$\mathcal{L}_{L1} = \mathbb{E}_{x,y,z}[\|y - G(x,z)\|_1] \quad (13)$$

L1 loss is chosen in this specific case as it encourages less blurring. This makes the final objective of the GAN to be:

$$G^* = arg \min_G \max_D \mathcal{L}_{cGAN}(G,D) + \lambda \mathcal{L}_{L1}(G) \quad (14)$$

where L1 loss is mixed with CGAN loss with a weight of $\lambda$.

The first layer of the discriminator is a Conv4 layer. Afterwards, there are three blocks containing a dense, a Conv4, and a normalization layer. The last layer is a Conv4 layer preceded by a dense layer with a leaky ReLU activation function.

*3.4 Evaluation Metrics*

*3.4.1 Evaluation of extracted fECG signal*

The recorded fECG signal from the fetus' scalp is used as ground truth in both datasets. The datasets were split using the 5-Fold cross-validation technique. Then, to assess the quality of the extracted fECG, mean absolute error (MAE), root mean squared error (RMSE), Pearson's correlation coefficient (PCC), spectral correlation ($\eta_{spec}$), spectral RMSE ($RMSE_{spec}$) metrics were used. For a qualitative evaluation of the signal extraction performance, predicted and ground truth signals were plotted and visually compared. The following formulae are used to determine MAE, RMSE, PCC, spectral correlation and spectral RMSE:

$$MAE = \sum_{i=1}^{N} |x - x_i| \quad (15)$$

$$RMSE = \sqrt{\sum_{i=1}^{N}(x - x_i)^2} \quad (16)$$

$$PCC = \frac{\sum_{i=1}^{N}(x_i - \bar{x})(y_i - \bar{y})}{\sqrt{\sum_{i=1}^{N}(x_i - \bar{x})^2}\sqrt{\sum_{i=1}^{N}(y_i - \bar{y})^2}} \quad (17)$$

$$\eta_{spec} = 1 - \frac{1 - PCC(PSD(y_i), PSD(\bar{y}_i))}{PCC(PSD(y_i), PSD(x_i))} \quad (18)$$

$$RMSE_{spec} = \frac{RMS(PSD(y_i) - PSD(\bar{y}_i))}{RMS(PSD(\bar{y}_i))} \quad (19)$$

Here, $x$ is the signal of interest, $\bar{x}$ is the corresponding mean value, $RMS$ signifies the root mean square operation.

*3.4.2 Evaluation of fQRS detection*

A modified Engelse and Zeelenberg (EngZee) method (Engelse, 1979; Lourenço, 2012) was used to detect the fQRS following the proposed model's extraction of the fECG. Then, evaluation was conducted using the F1-Score as well as accuracy, precision and recall. The following formulae were used to determine the metrics:

$$Accuracy = \frac{TP+TN}{TP+TN+FP+FN} \quad (20)$$

$$Precision = \frac{TP}{TP+FP} \quad (21)$$

$$Recall = \frac{TP}{TP+FN} \qquad (22)$$

$$F1 = 2 \cdot \frac{Precision \cdot Recall}{Precision+Recall} \qquad (23)$$

where true positive, true negative, false positive, false negative are expressed by $TP, TN, FP, FN$, respectively.

*3.4.3 Evaluation of heart rate variability*

A healthy heart cyclically pumps blood with a heart rate almost constant, with minimal variability. In diseased states, heart rate rhythm is distorted and variability in heartbeats can be detected. Comparison of healthy and unhealthy heart rhythms, particularly heart rate variability will reveal arrhythmia. We have used several metrics to monitor the heart of the fetus suggested by Shaffer et al. (Shaffer, 2017).

*Mean R-R interval* ($\mu_{RR}$): This is the average time interval between two consecutive R peaks of the ECG signal, often measured in milliseconds (ms). If we define the number of R peaks by $N$ and the time of R-peak by $T_{RR}$, $\mu_{RR}$ can be defined by the following equation:

$$\mu_{RR} = \frac{\sum_{i=1}^{N-1} T_{RR}(i+1) - T_{RR}(i)}{N-1} \qquad (24)$$

*Mean heart rate* ($\mu_{HR}$): Heart rate is one of the most prominent metrics of heart monitoring which signifies the number of heart bits per minute.

*Standard deviation of heart rate* ($\sigma_{HR}$): The heart does not oscillate like a metronome; instead, the heart rate varies within a range. Standard deviation of heart rate is a measure of how much the heart rate deviates from the expected value. Generally, it is measured in bits per minute (bpm) and can be calculated from the following equation:

$$\sigma_{HR} = \frac{\sum_{i=1}^{N} HR(i) - \mu_{HR}}{N-1} \qquad (25)$$

*Root mean square of successive RR interval differences* ($RMSSD$): $RMSSD$ is one of the most popular metrics for HRV and can be calculated by measuring the summation of the squared RR interval differences and then taking root.

$$RMSSD = \sqrt{\frac{\sum_{i=1}^{N} \left(T_{RR}(i+1) - T_{RR}(i)\right)^2}{N-1}} \qquad (26)$$

*SDNN/RMSSD*: SDNN is the standard deviation of the normal R-R interval. This is a long-term evaluation of heart rate variability and correlates to low-frequency components of ECG. On the other hand, RMSSD corresponds to high-frequency components of ECG signal (Otzenberger, 1998). Hence, the ratio of SDNN and RMSSD signifies LF/HF frequency ratio to some extent and is very important in heart rate variability measurement.

*NN50* and *PNN50*: $NN50$ is the number of pairs of successive RR intervals with a time difference of more than 50 milliseconds. PNN50 is the proportion of the number of RR intervals varied by more than 50 ms and the total number of RR intervals.

$$PNN50 = \frac{NN50}{Total\ number\ of\ RR\ intervals} \qquad (27)$$

**4. Results**

In this section, we will discuss the performance of the proposed framework which will be divided into three subsections: fECG extraction, identifying fQRS and heart rate estimation. Later, we will describe ablation studies conducted to identify suitable hyperparameters and the computational complexity of the experiment.

*4.1 fECG extraction result*

The results for different combinations of three types of generators (Unet, Resnet and Self-FPN) and two types of discriminators (basic and Self-ONN based) are compiled in **Table 1**.

**Table 1**: fECG extraction performance on different models (Scores are average of five folds)

| Generator | Discriminator | PCC | Spec. Corr. | Spec. RMSE |
|---|---|---|---|---|
| Unet 256 | Basic | 0.78 | 0.815 | 0.529 |
| Unet 256 | Self | 0.77 | 0.816 | 0.529 |
| Unet 128 | Basic | 0.751 | 0.799 | 0.54 |
| Unet 128 | Self | 0.732 | 0.77 | 0.545 |
| Resnet 9 blocks | Basic | 0.864 | 0.872 | 0.472 |
| Resnet 9 blocks | Self | 0.855 | 0.869 | 0.481 |
| **Resnet 13 blocks** | **Basic** | **0.882** | **0.894** | **0.428** |
| Resnet 13 blocks | Self | 0.871 | 0.89 | 0.453 |
| Self-FPN | Basic | 0.762 | 0.805 | 0.504 |
| Self-FPN | Self | 0.738 | 0.779 | 0.522 |

From **Table 1**, it is evident that Resnet 13 blocks generator and Basic discriminator pair performed better with an average PCC score of 0.882. This score varied from 0.87 to 0.89 for the five folds depicted in **Table 2**, along with the other metrics.

**Table 2**: Fold-wise fECG extraction performance for the best model

| Fold | RMSE | MAE | PCC | Spec. Corr. | Spec. RMSE |
|---|---|---|---|---|---|
| 1 | 0.106 | 0.069 | 0.89 | 0.9 | 0.42 |
| 2 | 0.109 | 0.073 | 0.88 | 0.88 | 0.45 |
| 3 | 0.103 | 0.068 | 0.89 | 0.9 | 0.42 |
| 4 | 0.102 | 0.069 | 0.89 | 0.9 | 0.4 |
| 5 | 0.115 | 0.077 | 0.87 | 0.89 | 0.45 |

Predictions on three sample test images are shown with the ground truth in **Figure 9**. The extracted fECG signal is very similar to the ground truth. The R peaks are detected accurately and some morphological information is also preserved.

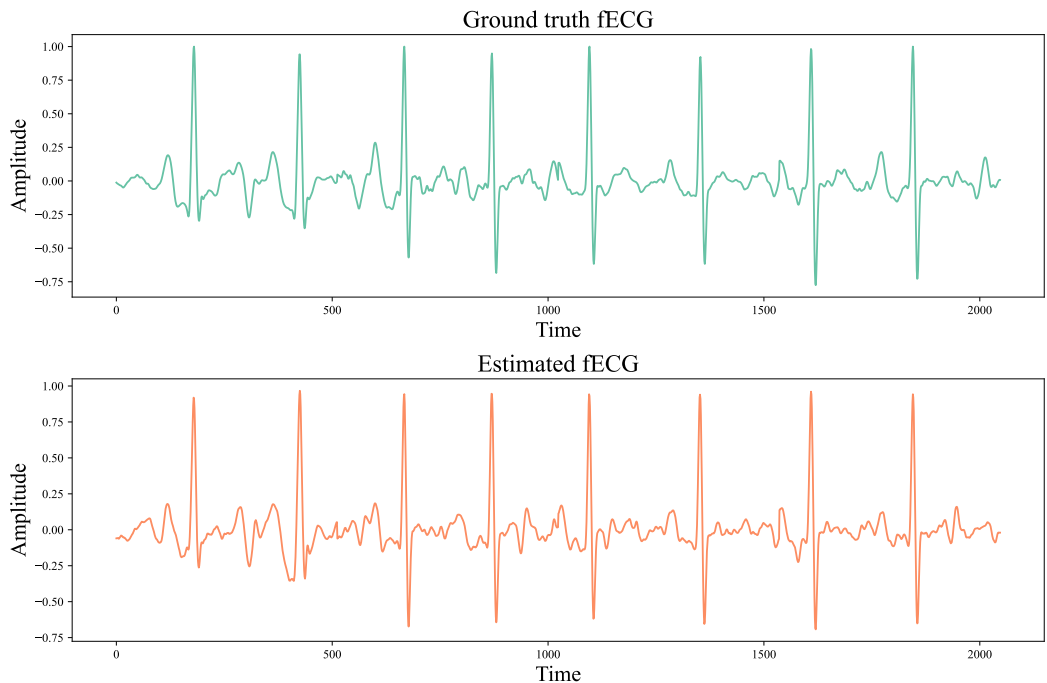

(a)

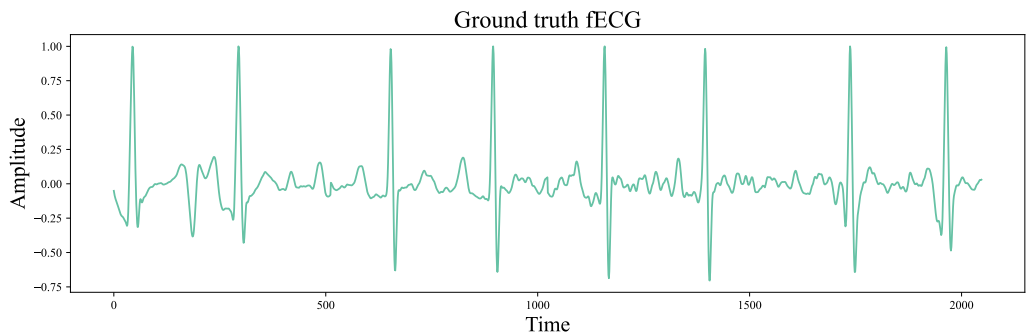
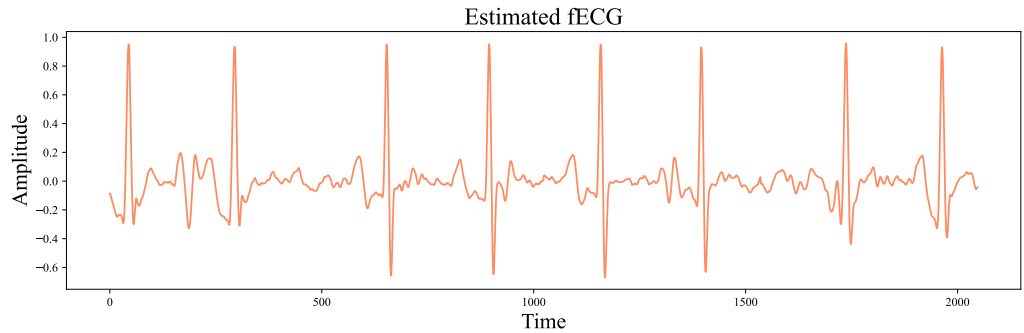

(b)

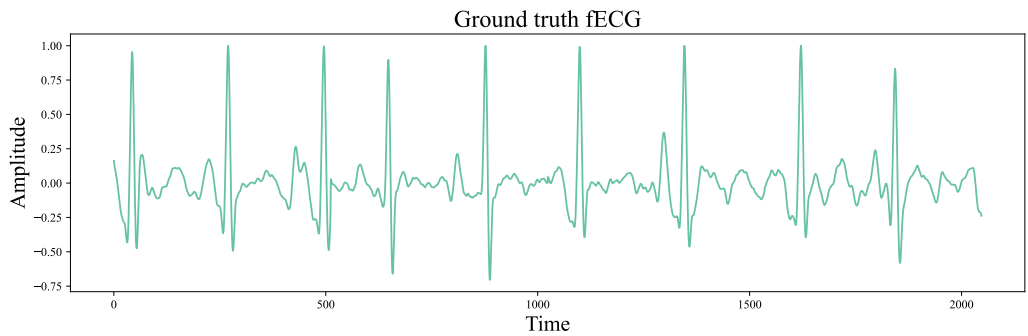
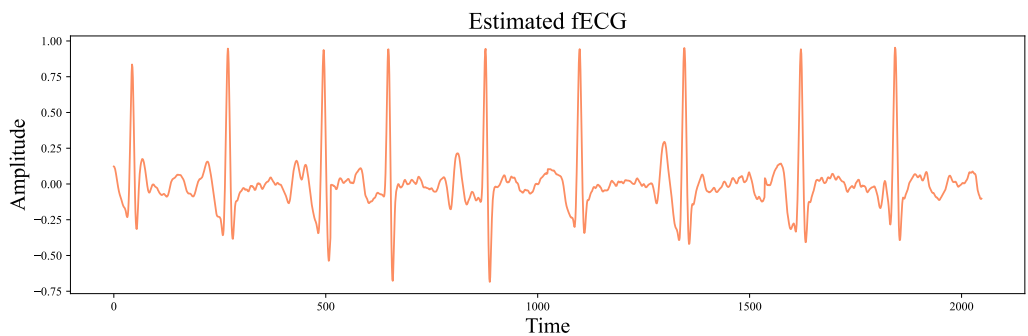

(c)

**Figure 9**: Three samples of extracted fECG signals using the best model (top: Ground Truth, bottom: Estimated fECG).

*4.2 fQRS detection result*

Once the fECG extraction is done, the fQRS is detected from the extracted ECG using the EngZee method proposed by Engelse and Zeelenberg. For measuring the performance of fQRS detection, any R-peak within 31.25 ms of the ground truth is considered to be correct. **Figure 10** contains the fQRS detection results on both ground truth and extracted fECG.

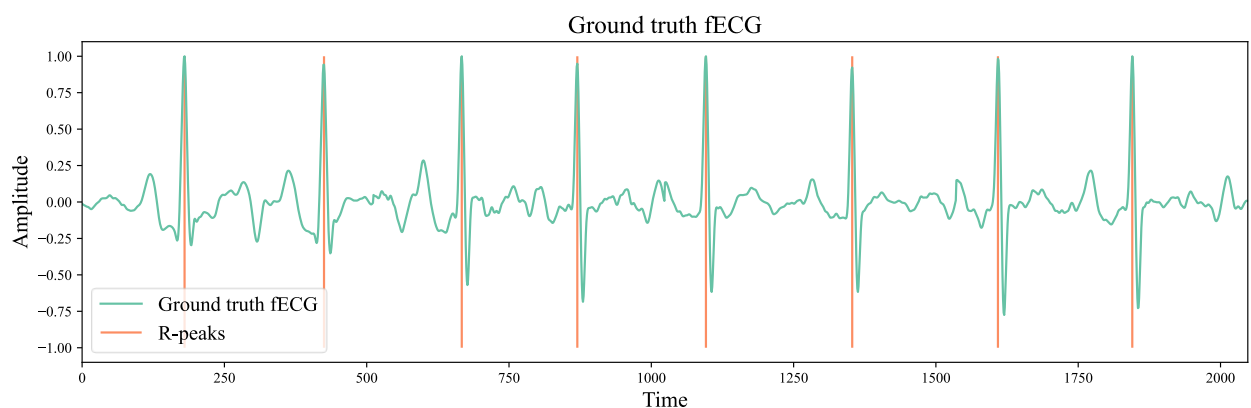
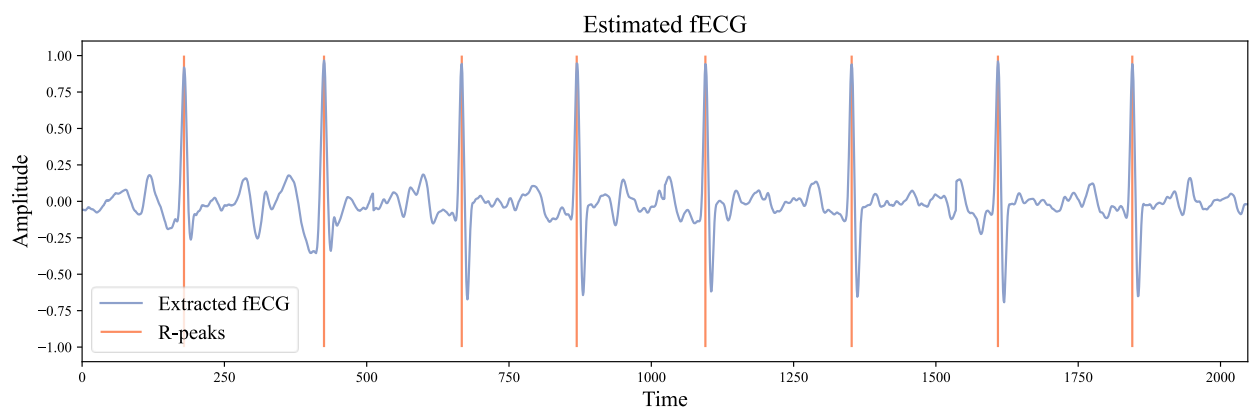

(a)

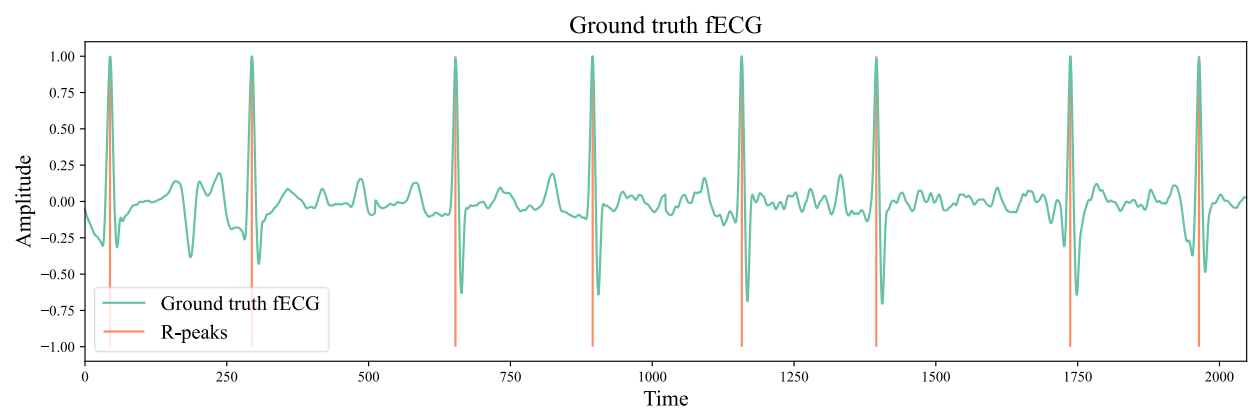
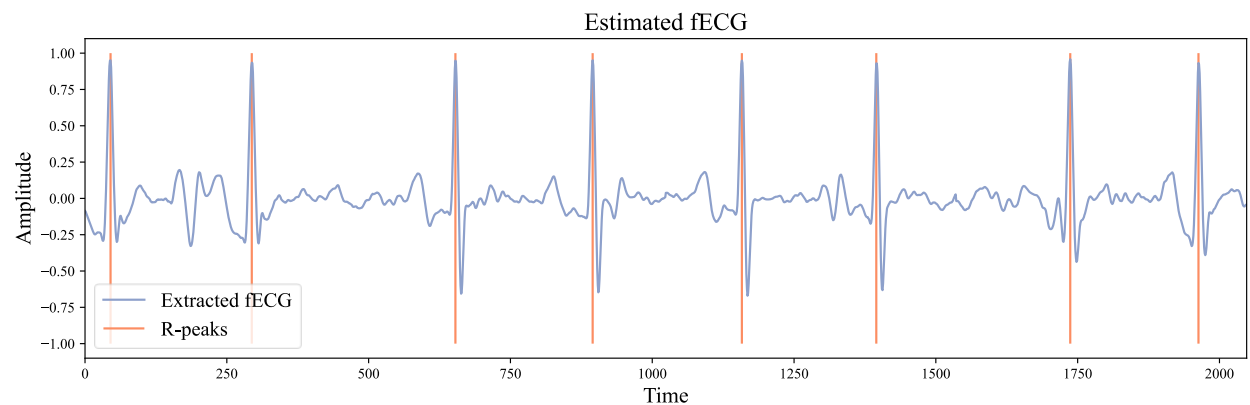

(b)

**Figure 10**: R-peak detection from ground truth and extracted fECG.

**Table 3** shows the related metrics for fQRS detection. All the metrics, most notably the F1 score, are uniform for all the folds, which signifies the robustness of the performance of our proposed methodology.

Table 3: fQRS Detection Performances on Test Set

| Fold | Precision | Recall | F1 Score | Accuracy |
|---|---|---|---|---|
| 1 | 0.98 | 0.95 | 0.96 | 0.93 |
| 2 | 0.97 | 0.95 | 0.96 | 0.92 |
| 3 | 0.98 | 0.95 | 0.97 | 0.93 |
| 4 | 0.97 | 0.95 | 0.96 | 0.92 |
| 5 | 0.98 | 0.94 | 0.97 | 0.92 |
| **Average** | 0.976 | 0.948 | 0.964 | 0.926 |

We present a brief comparison of our proposed technique with some other proposed methods in **Table 4**. Although Zhang et al. achieved a 99.4% F1 score with only smooth window SVD and adaptive filter, their study was restricted to only two persons in two separate tests. Mohebbian et al. also achieved excellent performance with an F1 score of 99.7%, but their methodology concentrates primarily on fQRS detection, ignoring the morphology of the ECG signal. Castillo et al. obtained an F1 score of 98.63% only on the selected signals by the medical specialists. However, considering both good and bad signals, their F1 score comes down to 94.11%.

Table 4: Comparison with other studies

| fECG extraction method | QRS detection method | Dataset | F1 Score (%) |
|---|---|---|---|
| SWSVD and adaptive filter (N. Zhang, 2017) | Pan-Tompkins (PT) | 1 | 99.4 |
| Compressive sensing (Da Poian, 2017) | Thresholding | 1 | 92.2 |
| ICA on compressed signal (Gurve, 2017) | Thresholding | 1 | 92.5 |
| Wavelet-based signal denoising (Castillo, 2018) | Clustering | 1 | 98.63 |
| Non-Negative Matrix Factorization (Gurve, 2019) | PT | 1 | 94.8 |
| Residual Encoder-Decoder network (Zhong, 2019) | PT | 1 | 94.1 |
| ICA, RLS, CWT (Jaros, 2019) | CWT | 2 | 90.99 |
| Principle Component Analysis (PCA) (Y. Zhang, 2020) | Clustering | 1 | 96.09 |
| ICA, RLS, EMD (Barnova, 2020) | WT | 2 | 90.1 |
| ICA, Fast Transversal Filter (FTF), Complementary EEMD (Barnova, 2021a) | Continuous Wavelet Transform (CWT) | 2 | 95.86 |
| Ensemble Empirical Mode Decomposition (EEMD) (Barnova, 2021b) | CWT | 2 | 95.69 |
| Attention-based CycleGAN (Mohebbian, 2021) | PT | 1 | 99.70 |
| STFT and GAN (Zhong, 2021) | PT | 2 | 90.05 |
| **Proposed 1D CycleGAN** | **EngZee** | **1 & 2** | **96.4** |

*4.3 Heart rate variability estimation*

One of the most useful applications of fECG is estimating fetal heart rate, which indicates the condition of the fetal heart and can identify several abnormalities. Additionally, we have estimated other metrics, including $\mu_{RR}, \mu_{HR}, \sigma_{HR}, RMSSD$ and $PNN50$. The metrics shown in **Table 5** indicate the performance of the proposed methodology.

Table 5: Heart rate metrics on Test Set

| Fold | ECG | $\mu_{RR}$ (ms) | $\mu_{HR}$ (bpm) | $\sigma_{HR}$ (bpm) | SDNN/RMSSD | PNN50 (%) |
|---|---|---|---|---|---|---|
| 1 | Ground truth | 473.19 | 128.52 | 14.14 | 0.7 | 40.45 |
|  | Extracted | 474.75 | 129.05 | 19.5 | 0.71 | 42.89 |
| 2 | Ground truth | 473.18 | 128.52 | 14.14 | 0.7 | 40.45 |
|  | Extracted | 476.91 | 128.23 | 18.32 | 0.69 | 42.62 |
| 3 | Ground truth | 473.19 | 128.51 | 14.15 | 0.71 | 40.45 |
|  | Extracted | 473.3 | 129.1 | 17.87 | 0.69 | 41.72 |
| 4 | Ground truth | 473.18 | 128.52 | 14.14 | 0.69 | 40.44 |
|   | Extracted    | 473.47  | 129.02  | 18.0   | 0.69  | 40.05  |
| 5 | Ground truth | 473.2   | 128.53  | 14.14  | 0.7   | 40.45  |
|   | Extracted    | 473.89  | 128.87  | 17.53  | 0.7   | 39.65  |
| **Average** | **Ground truth** | 473.188 | 128.52  | 14.142 | 0.7   | 40.448 |
|   | **Extracted**    | 474.464 | 128.852 | 18.244 | 0.696 | 41.386 |
| **Error (%)** |  | **0.27** | **0.25** | **29** | **0.57** | **2.32** |

Except for the standard deviation of HR, all the metrics are reasonably accurate. Moreover, a stable reactive fetal heart rate is typically between 120 to 160 bpm with a variability greater than 6 bpm (Rochard, 1976). In our case, the mean fetal heart rate is close to 129±18 bpm, which is well within the normal heart rate range.

*4.4 Ablation studies*

Ablation studies bear great importance in experimental studies. In an ablation study, a component or a parameter is changed to evaluate the significance of that portion of the system. In this section, we perform an ablation study of our proposed methodology on our proposed loss and two other parameters, learning rate and the number of epochs.

*Proposed loss*: In our proposed methodology, we introduced a weighted adversarial loss consisting of MSE loss, spectral loss, temporal loss, and power loss. This loss is very effective, especially for ECG signals. According to **Figure 11(a)**, the proposed loss increased the PCC score of the extracted fECG to almost 10%.

*Learning rate*: The learning rate is a very crucial hyperparameter in model training. In our study, we used different learning rates to find the most optimized result. For the best model (Resnet 13 blocks + Basic discriminator), we tried three different learning rates: 0.0001, 0.00001 and 0.000001 with enough epochs to reach the maxima. A learning rate of 0.00001 works best compared to others as summarized in **Figure 11(b)**.

*Number of epochs*: Generally, tuning the number of epochs can result in better fitness. In this study, we varied the number of epochs from 100 to 300 and evaluated the extracted fECG on the validation set. 150 epochs give better results than others, while 100 epochs seem to underfit, and 300 epochs seem to be overfitting as the PCC score decreases significantly, as shown in **Figure 11(c)**.

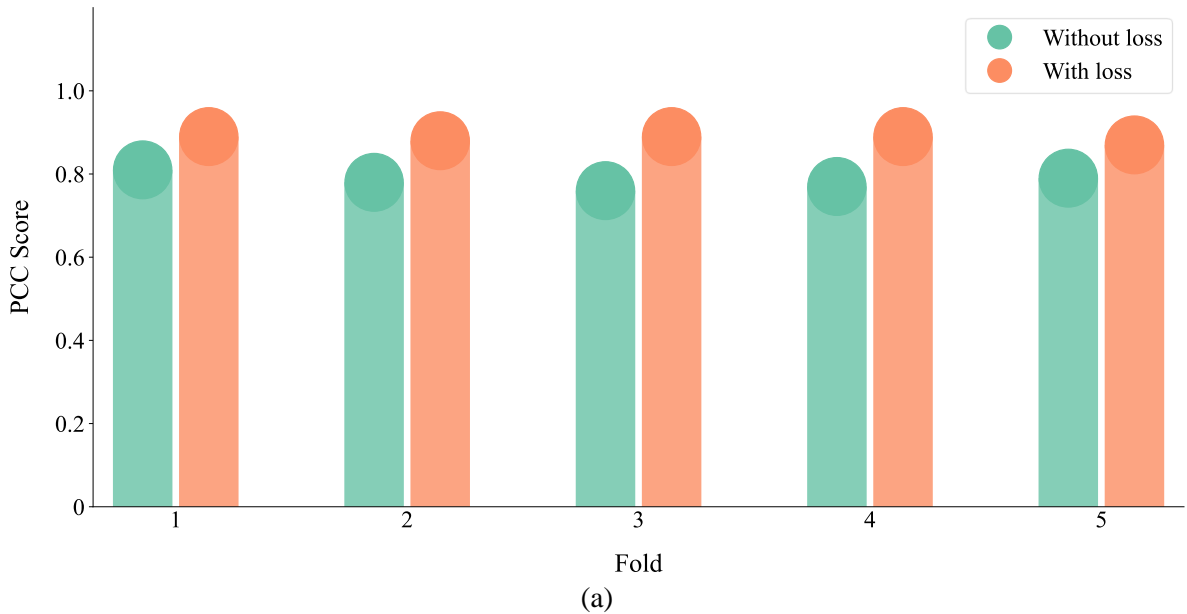

(a)

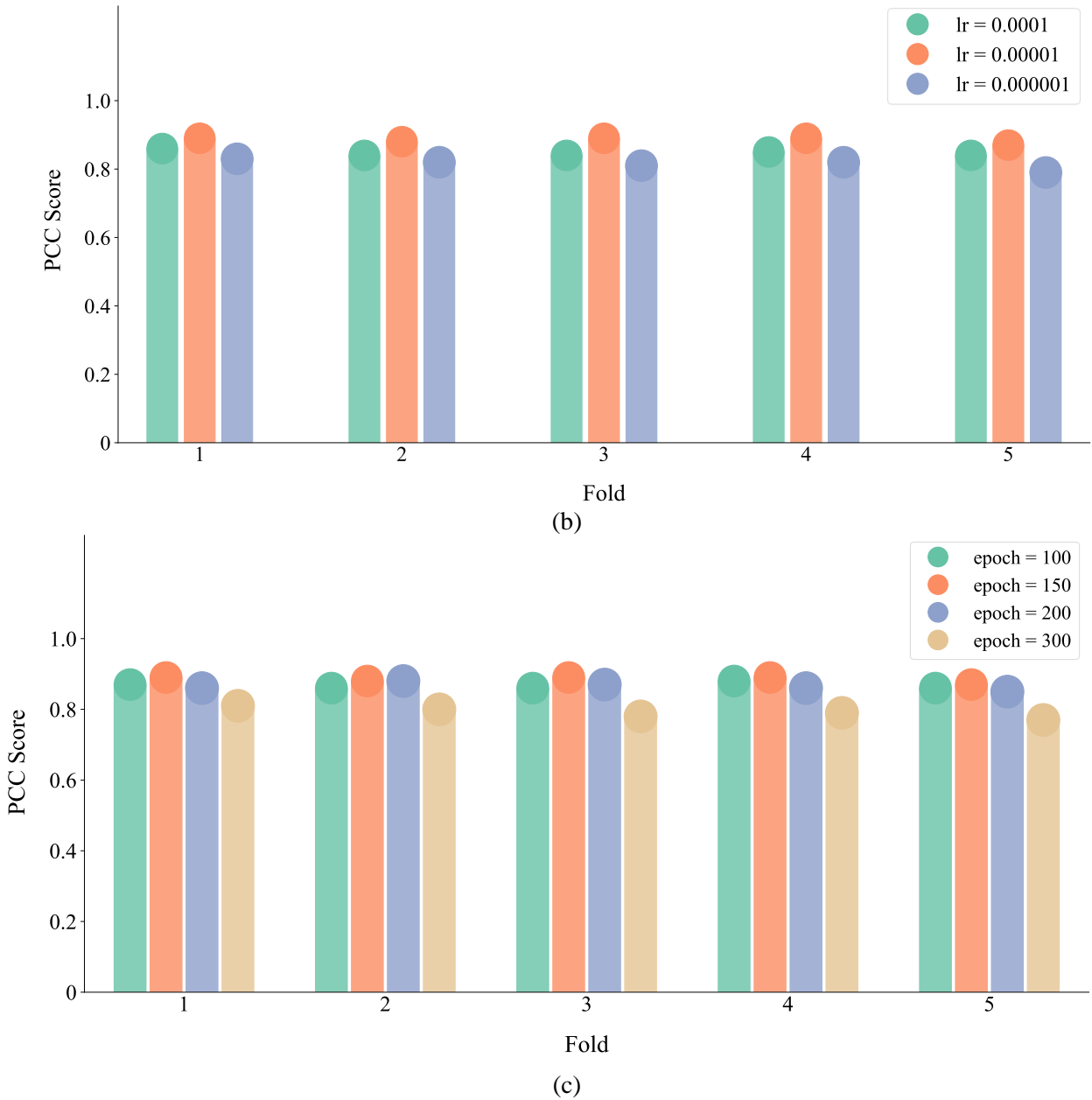

**Figure 11**: Ablation study of the proposed loss function (a), different learning rates (b), and the number of training epochs (c).

*4.5 Computational complexity*

For computational complexity analysis, we will mention the experimental computation setup along with the complexity of our proposed model. The experiments were conducted on a cloud virtual machine consisting of a 4-core 8-thread 2.0 GHz Intel Xeon processor with 25 GB of RAM and 16 GB NVIDIA Tesla T4 graphics card having a base frequency of 585 MHz We used Python 3.8 as the programming language and PyTorch, Numpy and Sci-kit Learn libraries for implementing network architecture, computation and evaluation, respectively. For the complexity of our proposed model, there are 0.337 million trainable parameters in each of the generators and 0.044 million in each of the discriminators making a total of 0.762 million parameters for the whole network. On the experimental setup, we extracted 1740 seconds long fECG from 4-channel mECG in only about 73 seconds. It made our proposed system suitable for real-time fECG extraction.

**5. Conclusion**

In this study, we have reconstructed Fetal ECG (fECG) from Mother ECG (mECG) using 1D CycleGAN and have measured heart rate and heart rate variability from the reconstructed signal. Traditional filtering methods are ineffective because mECG is frequently affected by baseline wandering, motion artifacts, power-line noise, uterine and muscle contraction, electrode connection, white noise etc. Our solution achieves a comparable performance to state-of-the-art techniques by showing excellent retainment of signal morphology with an average PCC and Spectral-Correlation score of 88.4% and 89.4%, respectively. We can detect fQRS of the signal with accuracy, precision, recall and F1 score of 92.6%, 97.6%, 94.8% and 96.4%, respectively. Moreover, the generated signal can estimate heart rate and R-R interval with 0.25% and 0.27% error, respectively. The main contributions of our work are excellent retention of components of the whole signal, detection of fQRS with great accuracy and near-perfect determination of fetal heart rate. It can be used to reconstruct fECG signals from any mECG signals for non-invasive fetal cardiac diagnosis and heart-rate measurements. Although the proposed framework can estimate heart rate with very high accuracy, the performance could be enhanced if the mECG signals were of a higher quality. One possible solution to this would be to create a dataset with similar protocols but with multiple expert annotations for ground truth fQRS positions (Modi, 2011). Furthermore, in the future, we can utilize another binary classification model that can select correct or discard excessively noisy mECG signals. In addition, several post-processing techniques, such as a refinement network, can be used to improve the fECG output quality further.

**Acknowledgment**
This study was funded by Qatar University Grant: QUHI-CENG-22/23-548. The open-access publication of this article was funded by the Qatar National Library.

**References:**
Anisha, M., Kumar, S., Nithila, E. E., & Benisha, M. (2021). Detection of Fetal Cardiac Anomaly from Composite Abdominal Electrocardiogram. *Biomedical Signal Processing and Control, 65*, 102308.
Bailey, J. J., Berson, A. S., Garson Jr, A., Horan, L. G., Macfarlane, P. W., Mortara, D. W., & Zywietz, C. (1990). Recommendations for standardization and specifications in automated electrocardiography: bandwidth and digital signal processing. A report for health professionals by an ad hoc writing group of the Committee on Electrocardiography and Cardiac Electrophysiology of the Council on Clinical Cardiology, American Heart Association. *Circulation, 81*(2), 730-739.
Barnova, K., Martinek, R., Jaros, R., & Kahankova, R. (2020). Hybrid methods based on empirical mode decomposition for non-invasive fetal heart rate monitoring. *IEEE Access, 8*, 51200-51218.
Barnova, K., Martinek, R., Jaros, R., Kahankova, R., Behbehani, K., & Snasel, V. (2021a). System for adaptive extraction of non-invasive fetal electrocardiogram. *Applied Soft Computing, 113*, 107940.
Barnova, K., Martinek, R., Jaros, R., Kahankova, R., Matonia, A., Jezewski, M., . . . Jezewski, J. (2021b). A novel algorithm based on ensemble empirical mode decomposition for non-invasive fetal ECG extraction. *PloS one, 16*(8), e0256154.
Behar, J., Johnson, A., Clifford, G. D., & Oster, J. (2014). A comparison of single channel fetal ECG extraction methods. *Annals of biomedical engineering, 42*(6), 1340-1353.
Castillo, E., Morales, D. P., García, A., Parrilla, L., Ruiz, V. U., & Álvarez-Bermejo, J. A. (2018). A clustering-based method for single-channel fetal heart rate monitoring. *PloS one, 13*(6), e0199308.
Chen, Z., Zeng, Z., Shen, H., Zheng, X., Dai, P., & Ouyang, P. (2020). DN-GAN: Denoising generative adversarial networks for speckle noise reduction in optical coherence tomography images. *Biomedical Signal Processing and Control, 55*, 101632.
Chu, C., Zhmoginov, A., & Sandler, M. (2017). Cyclegan, a master of steganography. *arXiv preprint arXiv:1712.02950*.
Clifford, G. D., Azuaje, F., & Mcsharry, P. (2006). ECG statistics, noise, artifacts, and missing data. *Advanced methods and tools for ECG data analysis, 6*(1), 18.
Clifford, G. D., Silva, I., Behar, J., & Moody, G. B. (2014). Non-invasive fetal ECG analysis. *Physiological measurement, 35*(8), 1521.
Da Poian, G., Rozell, C. J., Bernardini, R., Rinaldo, R., & Clifford, G. D. (2017). Matched filtering for heart rate estimation on compressive sensing ECG measurements. *IEEE Transactions on Biomedical Engineering, 65*(6), 1349-1358.
Engelse, W. A., & Zeelenberg, C. (1979). A single scan algorithm for QRS-detection and feature extraction. *Computers in cardiology, 6*(1979), 37-42.
Ferrara, E. R., & Widraw, B. (1982). Fetal electrocardiogram enhancement by time-sequenced adaptive filtering. *IEEE Transactions on Biomedical Engineering*(6), 458-460.
Goldberger, A. L., Amaral, L. A., Glass, L., Hausdorff, J. M., Ivanov, P. C., Mark, R. G., . . . Stanley, H. E. (2000). PhysioBank, PhysioToolkit, and PhysioNet: components of a new research resource for complex physiologic signals. *Circulation, 101*(23), e215-e220.

Goodfellow, I., Pouget-Abadie, J., Mirza, M., Xu, B., Warde-Farley, D., Ozair, S., . . . Bengio, Y. (2020). Generative adversarial networks. *Communications of the ACM, 63*(11), 139-144.

Gurve, D., & Krishnan, S. (2019). Separation of fetal-ECG from single-channel abdominal ECG using activation scaled non-negative matrix factorization. *IEEE Journal of Biomedical and Health Informatics, 24*(3), 669-680.

Gurve, D., Pant, J. K., & Krishnan, S. (2017). *Real-time fetal ECG extraction from multichannel abdominal ECG using compressive sensing and ICA.* Paper presented at the 2017 39th Annual International Conference of the IEEE Engineering in Medicine and Biology Society (EMBC).

Hasan, M., Ibrahimy, M., & Reaz, M. (2007). Techniques of FECG signal analysis: detection and processing for fetal monitoring. *WIT Transactions on Biomedicine and Health, 12*, 295-305.

He, K., Zhang, X., Ren, S., & Sun, J. (2016). *Deep residual learning for image recognition.* Paper presented at the Proceedings of the IEEE conference on computer vision and pattern recognition.

Hossain, M. S., Reaz, M. B. I., Chowdhury, M. E., Ali, S. H., Bakar, A. A. A., Kiranyaz, S., . . . Habib, R. (2022). Motion Artifacts Correction from EEG and fNIRS Signals using Novel Multiresolution Analysis. *IEEE Access, 10*, 29760-29777.

Isola, P., Zhu, J.-Y., Zhou, T., & Efros, A. A. (2017). *Image-to-image translation with conditional adversarial networks.* Paper presented at the Proceedings of the IEEE conference on computer vision and pattern recognition.

Jaros, R., Martinek, R., Kahankova, R., & Koziorek, J. (2019). Novel hybrid extraction systems for fetal heart rate variability monitoring based on non-invasive fetal electrocardiogram. *IEEE Access, 7*, 131758-131784.

Jeffries, P. R., Woolf, S., & Linde, B. (2003). Technology-based vs. traditional instruction: A comparison of two methodsfor teaching the skill of performing a 12-lead ecg. *Nursing education perspectives, 24*(2), 70-74.

Jezewski, J., Matonia, A., Kupka, T., Roj, D., & Czabanski, R. (2012). Determination of fetal heart rate from abdominal signals: evaluation of beat-to-beat accuracy in relation to the direct fetal electrocardiogram. *Biomedizinische Technik/Biomedical Engineering, 57*(5), 383-394.

Kaneko, T., Kameoka, H., Tanaka, K., & Hojo, N. (2019). *Cyclegan-vc2: Improved cyclegan-based non-parallel voice conversion.* Paper presented at the ICASSP 2019-2019 IEEE International Conference on Acoustics, Speech and Signal Processing (ICASSP).

Kiranyaz, S., Ince, T., Chowdhury, M. E., Degerli, A., & Gabbouj, M. (2022). Biosignal time-series analysis. In *Deep Learning for Robot Perception and Cognition* (pp. 491-539): Elsevier.

Lourenço, A., Silva, H., Leite, P., Lourenço, R., & Fred, A. L. (2012). *Real Time Electrocardiogram Segmentation for Finger based ECG Biometrics.* Paper presented at the Biosignals.

Martinek, R., Kahankova, R., Jezewski, J., Jaros, R., Mohylova, J., Fajkus, M., . . . Nazeran, H. (2018). Comparative effectiveness of ICA and PCA in extraction of fetal ECG from abdominal signals: Toward non-invasive fetal monitoring. *Frontiers in physiology, 9*, 648.

Matonia, A., Jezewski, J., Kupka, T., Jezewski, M., Horoba, K., Wrobel, J., . . . Kahankowa, R. (2020). Fetal electrocardiograms, direct and abdominal with reference heartbeat annotations. *Scientific data, 7*(1), 1-14.

Modi, S., Lin, Y., Cheng, L., Yang, G., Liu, L., & Zhang, W. (2011). A socially inspired framework for human state inference using expert opinion integration. *IEEE/ASME Transactions on Mechatronics, 16*(5), 874-878.

Mohebbian, M. R., Alam, M. W., Wahid, K. A., & Dinh, A. (2020). Single channel high noise level ECG deconvolution using optimized blind adaptive filtering and fixed-point convolution kernel compensation. *Biomedical Signal Processing and Control, 57*, 101673.

Mohebbian, M. R., Vedaei, S. S., Wahid, K. A., Dinh, A., Marateb, H. R., & Tavakolian, K. (2021). Fetal ECG extraction from maternal ECG using attention-based CycleGAN. *IEEE Journal of Biomedical and Health Informatics, 26*(2), 515-526.

Mumford, D., & Desolneux, A. (2010). *Pattern theory: the stochastic analysis of real-world signals*: CRC Press.

Niknazar, M., Rivet, B., & Jutten, C. (2012). Fetal ECG extraction by extended state Kalman filtering based on single-channel recordings. *IEEE Transactions on Biomedical Engineering, 60*(5), 1345-1352.

Otzenberger, H., Gronfier, C., Simon, C., Charloux, A., Ehrhart, J., Piquard, F., & Brandenberger, G. (1998). Dynamic heart rate variability: a tool for exploring sympathovagal balance continuously during sleep in men. *American Journal of Physiology-Heart and Circulatory Physiology, 275*(3), H946-H950.

Peters, M., Crowe, J., Piéri, J.-F., Quartero, H., Hayes-Gill, B., James, D., . . . Shakespeare, S. (2001). Monitoring the fetal heart non-invasively: a review of methods.

Rafaely, B., & Elliot, S. J. (2000). A computationally efficient frequency-domain LMS algorithm with constraints on the adaptive filter. *IEEE Transactions on Signal Processing, 48*(6), 1649-1655.
Rochard, F., Schifrin, B. S., Goupil, F., Legrand, H., Blottiere, J., & Sureau, C. (1976). Nonstressed fetal heart rate monitoring in the antepartum period. *American Journal of Obstetrics and Gynecology, 126*(6), 699-706.
Sameni, R., & Clifford, G. D. (2010). A review of fetal ECG signal processing; issues and promising directions. *The open pacing, electrophysiology & therapy journal, 3*, 4.
Shaffer, F., & Ginsberg, J. (2017). An overview of heart rate variability metrics and norms. Front Public Health. 2017; 5: 258. In: Epub 2017/10/17. https://doi. org/10.3389/fpubh. 2017.00258 PMID: 29034226.
Shepoval'nikov, R., Nemirko, A., Kalinichenko, A., & Abramchenko, V. (2006). Investigation of time, amplitude, and frequency parameters of a direct fetal ECG signal during labor and delivery. *Pattern Recognition and Image Analysis, 16*(1), 74-76.
Shuzan, M. N. I., Chowdhury, M. H., Hossain, M. S., Chowdhury, M. E., Reaz, M. B. I., Uddin, M. M., . . . Ali, S. H. M. (2021). A novel non-invasive estimation of respiration rate from motion corrupted photoplethysmograph signal using machine learning model. *IEEE Access, 9*, 96775-96790.
Tmenova, O., Martin, R., & Duong, L. (2019). CycleGAN for style transfer in X-ray angiography. *International journal of computer assisted radiology and surgery, 14*, 1785-1794.
Tran, L. D., Nguyen, S. M., & Arai, M. (2020). *GAN-based noise model for denoising real images.* Paper presented at the Proceedings of the Asian Conference on Computer Vision.
Varanini, M., Tartarisco, G., Balocchi, R., Macerata, A., Pioggia, G., & Billeci, L. (2017). A new method for QRS complex detection in multichannel ECG: Application to self-monitoring of fetal health. *Computers in biology and medicine, 85*, 125-134.
Yang, H., Sun, J., Carass, A., Zhao, C., Lee, J., Xu, Z., & Prince, J. (2018). *Unpaired brain MR-to-CT synthesis using a structure-constrained CycleGAN.* Paper presented at the Deep Learning in Medical Image Analysis and Multimodal Learning for Clinical Decision Support: 4th International Workshop, DLMIA 2018, and 8th International Workshop, ML-CDS 2018, Held in Conjunction with MICCAI 2018, Granada, Spain, September 20, 2018, Proceedings 4.
Zhang, N., Zhang, J., Li, H., Mumini, O. O., Samuel, O. W., Ivanov, K., & Wang, L. (2017). A novel technique for fetal ECG extraction using single-channel abdominal recording. *Sensors, 17*(3), 457.
Zhang, W., Yang, G., Lin, Y., Ji, C., & Gupta, M. M. (2018). *On definition of deep learning.* Paper presented at the 2018 World automation congress (WAC).
Zhang, Y., & Yu, S. (2020). Single-lead noninvasive fetal ECG extraction by means of combining clustering and principal components analysis. *Medical & Biological Engineering & Computing, 58*(2), 419-432.
Zhong, W., Liao, L., Guo, X., & Wang, G. (2018). A deep learning approach for fetal QRS complex detection. *Physiological measurement, 39*(4), 045004.
Zhong, W., Liao, L., Guo, X., & Wang, G. (2019). Fetal electrocardiography extraction with residual convolutional encoder–decoder networks. *Australasian physical & engineering sciences in medicine, 42*(4), 1081-1089.
Zhong, W., & Zhao, W. (2021). Fetal ECG extraction using short time Fourier transform and generative adversarial networks. *Physiological measurement, 42*(10), 105011.
Zhou, T., Krahenbuhl, P., Aubry, M., Huang, Q., & Efros, A. A. (2016). *Learning dense correspondence via 3d-guided cycle consistency.* Paper presented at the Proceedings of the IEEE Conference on Computer Vision and Pattern Recognition.
Zhu, J.-Y., Park, T., Isola, P., & Efros, A. A. (2017). *Unpaired image-to-image translation using cycle-consistent adversarial networks.* Paper presented at the Proceedings of the IEEE international conference on computer vision.